\documentstyle[aps,pra,prabib,amsfonts,amssymb,epsf]{revtex}
\twocolumn
\newcommand{\lkl}{\left(}
\newcommand{\rkl}{\right)}

\newcommand{\omegarf}{\bar{\omega}_{\rm{rf}}}
\newcommand{\omegaone}{\omega_{\rm{rf},1}}
\newcommand{\omegatwo}{\omega_{\rm{rf},2}}
\newcommand{\omegatrap}{\omega_{\rm{z}}}

\newcommand{\zres} {z_{\rm{rf},1,2}}
\newcommand{\zone}{z_{\rm{rf},1}}
\newcommand{\ztwo}{z_{\rm{rf},2}}

\newcommand{\deltarf}{\delta_{\rm{rf}}}

\newcommand{\Ai}{{\rm Ai}}
\newcommand{\Bi}{{\rm Bi}}

\begin{document}

\draft

% \markboth{\protect\Now}{\protect\Now}

\title{Investigations of a two-mode atom laser model}
\author{Jens~Schneider\thanks{E-mail: jens.schneider@mpq.mpg.de} and
  Axel~Schenzle\thanks{E-mail: axel.schenzle@physik.uni-muenchen.de}}

\address{Max-Planck-Institut f\"ur Quantenoptik,
  Hans-Kopfermann-Stra{\ss}e 1,
  D-85748~Garching and\\
  Sektion Physik, Ludwig-Maximilians-Universit\"at M\"unchen,
  Theresienstra\ss{}e 37, D-80333 M\"unchen, Germany}

%\date{submitted
%  to Phys. Rev. A, October 15, 1999}

\date{submitted to Phys.~Rev.~A, October 27, 1999}

% for timenow (\Now)}

\maketitle

\begin{abstract}
Atom lasers based on rf-outcoupling from a trapped Bose-Einstein
condensate can be described by a set of generalized, coupled
Gross-Pitaevskii equations (GPE). 

If not only one but two radio frequencies are used for outcoupling,
the atoms emerging from the trap have two different energies and the
total wavefunction of the untrapped spin-state is a coherent
superposition which leads to a pulsed atomic beam.

We present results for such a situation obtained from a 1D-GPE model
for magnetically trapped Rb-87 in the $F=1$ state. The wavefunction of
the atomic beam can be approximated by a sum of two Airy functions. In the
limit of weak coupling we calculate the intensity analytically.
\end{abstract}

\pacs{03.75.-Fi,05.30.Jp}

\narrowtext

\section{Introduction}
\label{sec:intro}

Now, as the creation of trapped Bose-Einstein condensates
\cite{ANDE95,BRAD95,DAVIS95} has nearly become a standard technique,
the interest in the field has turned to the manipulation of trapped
condensates. A major goal is the creation of coherent atomic beams
emerging from Bose condensates. Such a device, now called an atom
laser, was first realized in the MIT-group \cite{MEW97}. This
experiment created a rather small number of strong pulses of atoms.
Recently two groups were able to produce continuous
\cite{BLOCH99:1} or at least quasi-continuous \cite{HAGL99:1} coherent
atomic beams.

After the initial proposals of atom-laser models (see references in
\cite{PARK98:1}) the theoretical treatment of atom lasers has
focused on two major lines: the work in
\cite{BALL97,ZHAN98:1,ZHAN98:2,NARA97:2,STEC98:1,kneer98:_gener,%
HUTC99:1,JAPHA99:1,GRAH99:1,BAND99:1,EDWA99:1,SCHNEI99:1} starts
with the coupled Gross-Pitaevskii equations (GPE) (and generalizations
thereof like the Hartree-Fock-Bogoliubov theory) and analyzes the
corresponding solutions either numerically or analytically.

On the other hand, the studies in
\cite{HOPE97,MOY99:1,JACK99:1,HOPE99:1,hope99:_markov,BREU99:1}
concentrate on the characterization of atom lasers using master
equations analogous to the work on optical lasers and calculate
properties like the linewidth of atom lasers.

There are two different means of coherent outcoupling, {\em i.e}
getting atoms out of atom traps, which are investigated both
theoretically and experimentally. A radio frequency can be used to
flip the spin-state of magnetically trapped atoms to an untrapped
state; this so called radio-frequency outcoupling \cite{BALL97} has
been used in \cite{MEW97,BLOCH99:1}. An output coupler based on Raman
transitions was first described in \cite{MOY97} and implemented by the
NIST-group \cite{HAGL99:1}. In rf-outcoupling, the atomic beam just
falls downwards whereas two-photon Raman-outcoupling can be arranged
in such a way that the outcoupled atoms acquire momentum and the
resulting beam may thus be pointed in a specific direction.

Like in our previous work \cite{SCHNEI99:1} we concentrate on
rf-outcoupling describing the setup of the H\"ansch-group
\cite{BLOCH99:1} by a 1D-model based on coupled GPEs. In
\cite{SCHNEI99:1} we were able to characterize the experimental output
rate in a satisfactory way, differences were mainly due to the reduced
dimensionality of the model. Here, we want to investigate a slightly
different setup that was recently realized experimentally: instead of
only one rf-field Bloch {\em et al.} \cite{BLOCH99:2} used two radio
frequencies for outcoupling. Consequently, the coherently outcoupled
atoms have two different energies which leads to a pulsed, coherent
atomic beam. We will analyze the output properties of such a device
depending on the detunings of the radio frequencies and their
difference.

The paper has the following structure: In Sec.~\ref{sec:gpetheory} we
set the notation by introducing the basic facts about GPEs and rf-atom
lasers. In Sec.~\ref{sec:airy} we first show how stationary falling
waves can be described by appropriate combinations of Airy functions.
These are used to model the output of the two-mode atom laser and
to calculate its average output rate. The analytical results of this
section are then compared with numerical solutions of the coupled
GPEs (Sec.~\ref{sec:results}). It turns out that the visibility of the 
pulse pattern of the beam can be related to the correlation function of
the trapped Bose gas (here at $T=0$, see also \cite{BLOCH99:2}). In
Sec.~\ref{sec:concl} we summarize our results.

\section{Coupled Gross-Pitaevskii equations with two radio frequencies}
\label{sec:gpetheory}

The properties of Bose condensates of atomic vapors at zero
temperature are very well described by the Gross-Pitaevskii equation
(GPE) (see \cite{DALF98:1} and references therein). It is a nonlinear
Schr\"odinger equation for the order parameter of the system, namely
the macroscopic wavefunction of the Bose condensate, and accounts also
for the coherence properties of the condensate. It may thus serve as a
starting point for considerations in coherent atom optics with Bose
condensates as it was done in \cite{BALL97}: the authors generalized
the GPE to a system of coupled equations that describe the physics of
the trapped and untrapped states of a Bose gas coupled via an rf-field
including the interaction due to collisions.

In the present article, we modify this approach to a situation where
the trapped atoms are exposed to two radio frequencies. Such a setup
has already been analyzed experimentally \cite{BLOCH99:2}. We focus on
a trap with Rb-87 atoms in the $F=1$ hyperfine-manifold. The three
Zeeman sublevels with $m \in \{-1,0,1\}$ are described by macroscopic
wavefunctions $\psi_m$; after applying the dipole approximation the
system of generalized GPEs reads
\begin{eqnarray}
  \label{eq:gpe3dbare}
  i \hbar \frac{\partial}{\partial t} \psi_m(\vec r, t) &=& \nonumber \\
  && \hspace{-1.4cm} \Bigg(-\frac{\hbar^2\nabla^2}{2M} + V_m(\vec r) 
      + U ||\psi(\vec r, t)||^2 \Bigg) \psi_m(\vec r, t) \nonumber \\
  && \hspace{-1.4cm} + \sum_{m'} W_{m,m'} \psi_{m'}(\vec r, t)\;. 
\end{eqnarray}
We assume that all Zeeman levels interact with the same s-wave
scattering length $a_0 = 110\, a_{\rm Bohr}$ so $U = 4\pi\hbar^2 a_0
N/M$. $||\psi(\vec r, t)||^2 = \sum_m |\psi_m(\vec r, t)|^2$ denotes
the total density in the trap divided by the number $N$ of particles.
The interaction between the atoms and the radio frequency fields with
frequencies $\omegaone$ and $\omegatwo$ is described by
\begin{eqnarray}
  \label{eq:Wmm}
  W_{m,m'} &=& 2 \hbar\Omega
  \big(\cos(\omegaone t)+\cos(\omegatwo t)\big) \nonumber\\
  && \qquad\quad\times\big(\delta_{m,m'+1} + \delta_{m,m'-1}\big)\;,
\end{eqnarray}
where $\Omega = g_F \mu_{\rm Bohr} B_{\rm rf}/(\sqrt{2}\hbar)$ ($g_F = 1/2$)
is the Rabi frequency corresponding to the magnetic field strength
$B_{\rm rf}$.

We now define $\omegarf = (\omegaone+\omegatwo)/2$ to be the average
radio frequency and $\delta_{\rm{rf}} = \omegaone-\omegatwo$ as the
difference between the two driving fields. In order to get rid of the high
frequencies, we apply the transformation $\psi_m(t) \rightarrow e^{-i
  m \omegarf t} \psi_m(t)$ and skip all terms $\propto\exp{(-2 i m
  \omegarf t)}$ (rotating wave approximation). This turns
Eq.~(\ref{eq:gpe3dbare}) into
\begin{eqnarray}
  \label{eq:gpe3drwa}
  i \hbar \frac{\partial}{\partial t} \psi_m(\vec r, t) &=&
        \Bigg(-\frac{\hbar^2\nabla^2}{2M} + V_m(\vec r) 
        +m\hbar\omegarf \nonumber\\
     && \quad + U ||\psi(\vec r, t)||^2 \Bigg) \psi_m(\vec r, t)
        \nonumber\\
     && + \sum_{m'} \tilde{W}_{m,m'}(t) \psi_{m'}(\vec r, t)
\end{eqnarray}
with
\begin{equation}
  \label{eq:Wmm2}
  \tilde{W}_{m,m'}(t) = 2 \hbar\Omega\cos\big(\delta_{\rm{rf}}t/2\big)
              \big(\delta_{m,m'+1} + \delta_{m,m'-1}\big)\;.
\end{equation}
Equation~(\ref{eq:gpe3drwa}) represents a set of equations in
three dimensions that has to be solved numerically. This is
computationally very demanding.  In earlier work
\cite{BALL97,NARA97:2,STEC98:1,SCHNEI99:1} a one-dimensional (1D)
approach has proven to be quite successful.  Usually, the traps used in
experiments are elongated in the horizontal plane. Like in
\cite{SCHNEI99:1} we choose coordinates where the $z$-axis points {\em
  downwards}, the long axis of the trap is denoted by $y$, the short
horizontal one by $x$. Taking gravity into account, the total
effective potentials in $z$-direction are then given by
\begin{eqnarray}
  \label{eq:veff}
  V_{m,{\rm eff}}(z,t) &=& -m\,M \omega_z^2 z^2/2
  + m \hbar \Delta \nonumber\\
                       && \hspace{2cm} - Mgz + g_{\rm 1D} ||\psi(z,t)||^2\;,
\end{eqnarray}
where $\Delta= \hbar\omegarf - V_{\rm off}$ denotes the detuning from
the transitions at the trap center ($V_{\rm off} = -g_F \mu_{\rm Bohr}
B_{\rm off}$).  Due to gravity the output coupling proceeds mainly in
the vertical direction so it is natural to choose this direction as
the 1D axis for our model.

To obtain the effective interaction strength $g_{\rm 1D}$ in
Eq.~(\ref{eq:veff}) we require the chemical potential in Thomas-Fermi
approximation in the 1D model (for the trapped $m=-1$ state) to equal
the one of the full 3D situation. This leads to
\begin{equation}
  \label{eq:g1d}
  g_{\rm 1D} = \frac{2}{3} \lkl \frac{\omega_y}{\omega_z} \rkl ^{\frac{3}{5}}
  \lkl \frac{15 N a_0}{a_z} \rkl ^{\frac{3}{5}} \hbar\omega_z a_z
\end{equation}
with $a_z = \sqrt{\hbar/(M\omega_z)}$.

The 1D approximation of Eq.~(\ref{eq:gpe3drwa}) then reads
\begin{eqnarray}
  \label{eq:gpe1d}
  i \hbar \frac{\partial}{\partial t} \psi_m(z, t) &=&
  \lkl -\frac{\hbar^2}{2M}\frac{\partial^2}{\partial z^2}
       + V_{m,{\rm eff}}(z,t)\rkl \psi_m(z, t) \nonumber \\
  && + \sum_{m'} \tilde{W}_{m,m'}(t) \psi_{m'} (z, t)\;.
\end{eqnarray}

\begin{figure}[htbp]
  \begin{center}
% RbF1_2freq.0060.film2.gnu, i= 87
    \epsfxsize=0.47\textwidth
    \epsffile{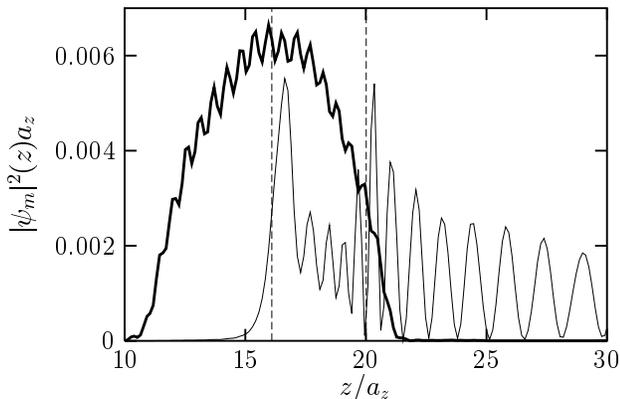}
    \caption{Plot of the densities for $m=-1,0$ ($F=1$) in the trap for 
      weak outcoupling. The trapped $m=-1$ state (thick line, scaled
      down by a factor of 20) shows some small excitations on top of
      the ground state. The untrapped $m=0$ state (thin line) exhibits 
      pulses for $z\gtrsim 23\,a_z$. The dashed lines denote the resonance
      points $\zres$ from Eq.~(\ref{eq:zres}), here we have
      $\omega_{{\rm rf}1,2} = 2\pi\times(21000 \pm 4500)\,\mbox{Hz}$.}
    \label{fig:densities2res}
  \end{center}
\end{figure}

In this work, we concentrate on the weak coupling regime of
Eq.~(\ref{eq:gpe1d}) which is characterized by $\Omega < \omega_z$
\cite{MEW97,STEC98:1,BAND99:1}. The condensate is hardly affected in
this regime, it merely changes its overall size due to the loss of
atoms by the outcoupling process \cite{STEC98:1}. In \cite{SCHNEI99:1}
we have pointed out that the coupling between the trapped and
untrapped states is concentrated spatially to the crossing points of
the effective potentials in Eq.~(\ref{eq:veff}). In the present case
where two radio frequencies are present, one expects two such points
of maximal coupling. They are located at
\begin{equation}
  \label{eq:zres}
  \zres = \sqrt{2(\Delta\pm\delta_{\rm rf}/2)
    / (\hbar\omega_z)}\, a_z\;,
\end{equation}
corresponding again to the crossing points of the effective potentials
for either the radio frequency $\omegaone$ or $\omegatwo$ (see
Fig.~\ref{fig:pots}).  In Fig.~\ref{fig:densities2res} the densities
of the states $m=-1$ and $0$ ($F=1$) are shown for a specific
situation. The condensate is in the $m=-1$ state, the plot shows a
snapshot of the density distributions after $5\,\mbox{ms}$ of
outcoupling (for further computational details see
Sec.~\ref{sec:results}). The distribution for the $m = 0$ state
demonstrates clearly that the points $\zres$ are indeed the points of
maximum outcoupling; if one considers the two radio frequencies
separately, these points correspond to the classical turning points in
the effective potentials in Eq.~(\ref{eq:veff}) (see
\cite{SCHNEI99:1}).
In the weak coupling region we are considering here, the state $m=1$,
which is also untrapped, is hardly populated at all. For this reason,
its density distribution is not shown in Fig.~\ref{fig:densities2res}
and we will not discuss it further. Nevertheless, it is included in
all numerical calculations.

\section{Airy functions for falling atoms and output flux}
\label{sec:airy}

\subsection{Stationary, falling waves}
\label{sec:fallwav}

Outside the condensate region, the untrapped $m=0$ state feels only
the gravitational potential because the mean field contribution in
$V_{0,{\rm eff}}$ is practically zero there. Therefore, the
wavefunction for the falling particles is a linear combination of Airy
functions, which are the solutions of the Schr\"odinger equation for a
linear potential. The usual Airy function $\Ai(\xi)$ does not suffice
to describe a continuous, stationary beam of falling atoms. In analogy
to free planar waves, where both cosine and sine are used to form a
traveling wave $\propto\exp(ikx)$, we have to use a special complex
solution of the Schr\"odinger equation for a given energy $E$
\begin{eqnarray}
  \label{eq:psiE}
  \psi_E(z,t) &=& \big\{\Ai(-\xi_E(z)) -i\Bi(-\xi_E(z))\big\}
  \,e^{-iE t/\hbar}  \nonumber\\
  &=& {\cal M}(\xi_E(z))\,e^{-i[\Theta(\xi_E(z))+Et/\hbar]}\;.
\end{eqnarray}
The argument $\xi_E(z)$ simply shifts and rescales the $z$-axis
\begin{eqnarray}
  \label{eq:xi}
  \xi_E(z) &=& \frac{1}{l}\lkl z + \frac{E}{Mg}\rkl, \qquad
  l = \lkl\frac{\hbar^2}{2M^2g}\rkl^{1/3}\;,
\end{eqnarray}
where l is a length scale (see e.g. \cite{FLUEGGE}). The wavefunction
$\psi_E(z,t)$, though being a solution of the Schr\"odinger equation
for a homogenous gravitational field, is not normalizable due to the
exponential growth of $\Bi(\xi_E(z))$ in the classical forbidden
region $z < -E/(Mg)$. This is to be expected, as it describes a
situation where particles are flowing steadily to $z\to\infty$ without
a source of particles at some finite $z$. Nevertheless, we are only
interested in the classically allowed region where the solution is
perfectly analogous to a freely falling wave.

We will later make use of the asymptotic forms ($z\to\infty$) of
${\cal M(\xi)}$ and $\Theta(\xi)$ \cite{ABRAMOWITZ}
\begin{eqnarray}
  \label{eq:asympt}
  {\cal M(\xi)} &\approx& \frac{1}{\sqrt{\pi}} \xi^{-1/4} \nonumber\\
  \Theta(\xi)   &\approx& \frac{\pi}{4} - \frac{2}{3} \xi^{3/2}
\end{eqnarray}
to obtain a nice interpretation for $|\psi_0(z,t)|^2$.

\subsection{Coherent superposition of two falling waves}
\label{sec:twowaves}

\begin{figure}[htbp]
  \begin{center}
    \epsfxsize=0.45\textwidth
    \epsffile{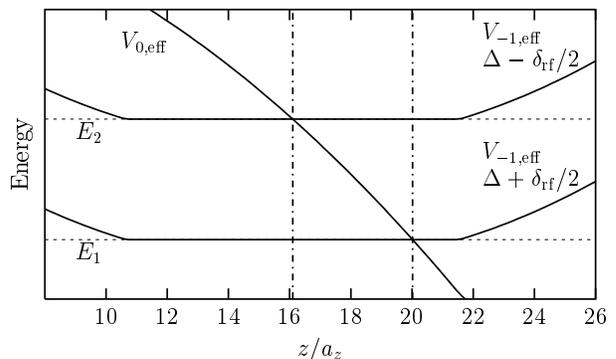}
    \caption{Effective potentials $V_{m,{\rm eff}}$ and energies
      $E_{1,2}$ of the outcoupled atoms for the two detunings $\Delta
      \pm \delta_{\rm rf}/2$. The two vertical lines denote the
      resonance points $\zres.$}
    \label{fig:pots}
  \end{center}
\end{figure}

The two radio frequencies driving the condensate should lead to two
different energies in the resulting atomic beam (see
Fig.~\ref{fig:pots}).
The energy of the outcoupled beam for a detuning $\Delta$ is 
given by
\begin{equation}
  \label{eq:energy}
  E = \mu - \frac{1}{2}\frac{Mg^2}{\omegatrap^2} - \hbar\Delta\;,
\end{equation}
where $\mu$ denotes the chemical potential.
If we consider the radio frequencies $\omegaone$ and $\omegatwo$, we
have to use the two detunings $\Delta_{1,2} =
\hbar\omega_{\rm{rf},1,2} - V_{\rm off}$ to get the energies
$E_{1,2}$ indicated in Fig.~\ref{fig:pots}.

Assuming that the outcoupling process proceeds in a fully coherent
way, the total wavefunction of the outcoupled beam  is given by
\begin{equation}
  \label{eq:psi0}
  \psi_0(z,t) = {\cal N}\big( \psi_{E_1}(z,t) + s \psi_{E_2}(z,t) \big)\;,
\end{equation}
where $\cal N$ is an overall normalization constant and $s$ accounts
for the relative strength of the two contributing waves.  In
experiment, one usually measures the probability density
$|\psi_0(z,t)|^2$ via some imaging technique. If the two energies
$E_{1,2}$ are not too different from each other with respect to the
mean energy $\bar{E} = (E_1+E_2)/2$, {\em i.e.} $\hbar\delta_{\rm
  rf}/{\bar E} \ll 1$, then Eq.~(\ref{eq:psi0}) can be Taylor-expanded
around $\bar E$.  Taking the asymptotic forms in Eq.~(\ref{eq:asympt})
into account, we can write $|\psi_0(z,t)|^2$ as
\begin{eqnarray}
  \label{eq:psi02}
  |\psi_0(z,t)|^2 &=& \tilde{{\cal N}}^2 {\cal M}^2(\xi_{\bar E}(z))
  \nonumber\\ 
  && \times \big\{ 2 + 2 P \cos (\phi(z,t)-\alpha) \big\}
\end{eqnarray}
with the phase
\begin{equation}
  \label{eq:phizt}
  \phi(z,t) = \delta_{\rm rf}t - \frac{\hbar\delta_{\rm rf}}{Mgl}
    \sqrt{\xi_{\bar E}(z)}\;,
\end{equation}
the modified norm
\begin{equation}
  \label{eq:modnorm}
  \tilde{{\cal N}}^2 = {\cal N}^2 \frac{1+|s|^2}{2}\;,
\end{equation}
and
\begin{equation}
  \label{eq:Pvons}
  P = \frac{2 |s|}{1+|s|^2}\;.
\end{equation}
Apart from the ${\cal M}^2$-dependence, Eq.~(\ref{eq:psi02}) has the
typical form of an interference term with $P$ being the visibility.
The maxima of this term describe the falling of pulses emerging
from the trap. They are located at
\begin{equation}
  \label{eq:zmax}
  z_k = -\frac{\bar E}{Mg} +
       \frac{1}{2}g \bigg( t - \frac{2\pi k + \alpha}{\delta_{\rm
           rf}}\bigg)^2,
       \qquad k\in{\mathbb{N}}_0\;,
\end{equation}
corresponding to a pulse frequency $\nu_{\rm pulse} = \delta_{\rm
  rf}/(2\pi)$. This is exactly in analogy to ``mode-locking'' of two
modes in a laser.

The phase $\alpha$ of $s = |s|e^{\alpha}$ appearing in the above
equations is responsible for a shift of the interference pattern i.e.
the pulses. Eq.~(\ref{eq:psi0}) describes the output of the atom laser
only after it has reached a kind of stationary operation. We therefore
use $\alpha$ merely as a free parameter to shift the pulses in time.

\subsection{Rate of outcoupling}
\label{sec:rates}

To determine the normalization factor of the wavefunction $\cal N$ in
Eq.~(\ref{eq:psi0}) we use the fact that in 1D the flux density $j_z$
equals directly the rate of atoms going through a point. $j_z$ can be
written as
\begin{eqnarray}
  \label{eq:jzprinc}
  j_z(z,t) &=& |\psi_0(z,t)|^2 v_z(z), \\
  \label{eq:vz}
  v_z(z)   &=& -\frac{\hbar}{M}\frac{\partial\Theta}{\partial z} =
  \frac{\hbar}{Ml}\sqrt{\xi_{\bar E}(z)}\;,
\end{eqnarray}
$v_z$ being the velocity of atoms in z-direction. Eq.~(\ref{eq:vz}) is
obtained by considering only the leading terms in $\xi_E(z)$ for
$z\to\infty$.
Averaging over time in Eq.~(\ref{eq:jzprinc}) finally leads to
\begin{equation}
  \label{eq:jz}
  j_z = \frac{\hbar}{2\pi Ml} \tilde{{\cal N}}^2\;.
\end{equation}

In \cite{STEC98:1}, Steck {\em et al.} gave a formula for the outcoupling
rate of a three dimensional trap in the weak coupling limit.
Analogously, this result can be derived for the 1D situation we are
interested in (see \cite{STEC97:1}). The rate for one resonance point
$z_{\rm res}$ (resulting from a detuning $\Delta)$ is given by
\begin{equation}
  \label{eq:gamma1d}
  \Gamma^{\rm 1d}_{z_{\rm res}} =
  2\pi\bigg(\frac{\Omega}{\omegatrap}\bigg)^2
  \frac{|\psi_{-1}(z_{\rm res})|^2}{\sqrt{2\Delta/\omegatrap}}a_z\omegatrap\;.
\end{equation}
We now assume that the rates at the two resonance points $\zres$ add
up to the total outcoupling rate. This rate must be equal to $j_z$, so
finally with the help of Eq.~(\ref{eq:jz}) we arrive at an expression
for the normalization
\begin{equation}
  \label{eq:norm}
  \tilde{{\cal N}}^2 = \frac{\pi Ml}{2\hbar}
             \lkl \Gamma^{\rm 1d}_{\zone} + \Gamma^{\rm 1d}_{\ztwo} \rkl\;,
\end{equation}
where $\Gamma^{\rm 1d}_{\zres}$ is obtained from the detunings
$\Delta\pm \deltarf /2$.

Inspection of Eq.~(\ref{eq:psi0}) shows that $|s|$ should
reflect the relative outcoupling rates at the two resonance points. We
define it via
\begin{equation}
  \label{eq:visib}
  |s|^2 = \frac{\Gamma^{\rm 1d}_{\zone}}
               {\Gamma^{\rm 1d}_{\ztwo}}\;,
\end{equation}
which leads to
\begin{equation}
  \label{eq:Pgamma}
  P = 2 \frac{\sqrt{\Gamma^{\rm 1d}_{\zone} \Gamma^{\rm 1d}_{\ztwo}}}
             {\Gamma^{\rm 1d}_{\zone} + \Gamma^{\rm 1d}_{\ztwo}}\;
\end{equation}
for the visibility.
The numerical results in the next section show that this is indeed
the correct expression. If the frequency difference $\delta_{\rm rf}$
is very small, one can also use
\begin{eqnarray}
  \label{eq:n2approx}
  \tilde{{\cal N}}^2 &=& \frac{\pi Ml}{\hbar}
  \Gamma^{\rm 1d}_{z_{{\rm res}}}\;,\\
  \label{eq:sapprox}
  |s|^2 &=& \frac{|\psi_{-1}(\zone)|^2}
               {|\psi_{-1}(\ztwo)|^2}\;,\\
  \label{eq:papprox}
  P &=& 2 \frac{|\psi_{-1}(\zone)| |\psi_{-1}(\ztwo)|}
               {|\psi_{-1}(\zone)| + |\psi_{-1}(\ztwo)|}\;,
\end{eqnarray}
which shows that $P$ is related to the correlation function of the
trapped Bose gas \cite{BLOCH99:2}.

\section{Numerical results}
\label{sec:results}

\subsection{Numerical method and observation of pulses}
\label{sec:method}

To solve the 1D GPEs in Eq.~(\ref{eq:gpe1d}) numerically we propagate
the wavefunctions with a usual split-operator algorithm using FFT on a
1D grid. As starting condition we always use the ground state of
the trapped $m=-1$ state, which we obtain by imaginary time
propagation.

The coupling of the Zeeman levels via the rf-field is implemented in a
very straightforward way: if the right hand side of
Eq.~(\ref{eq:gpe1d}) is written as $H\psi$, one can split up the
Hamiltonian $H$ in an obvious way $H=T+V(t)+W(t)$. The kinetic and
potential parts $T$ and $V(t)$ are taken care of by the
FFT-split-operator technique. It is sufficient to treat the coupling
part $W(t)$ separately in the most simple way by keeping only the
terms up to first order in the propagation time. The propagator from
time $t$ to $t+\Delta t$ then reads
\begin{equation}
  \label{eq:propa}
  U(t,t+\Delta t) = U_0(t,t+\Delta t)
                    \lkl 1 - \frac{i}{\hbar}W(t)\Delta t \rkl\;,
\end{equation}
where $U_0$ is the split-operator propagator for $T+V(t)$.

The trap parameters are basically taken from \cite{BLOCH99:1}, the
trapping frequency for the $m=1$ state is $\omega_z = 2\pi \times
127\,\mbox{Hz}$. We often use harmonic oscillator units, the length
unit is $a_z = 0.95\,\mbox{$\mu$m}$, time is measured in $1/\omega_z = 
1.3\,\mbox{ms}$. 

To investigate the output in our atom-laser model we have propagated
the initial wavefunction for up to $\approx 30\,\mbox{ms}$ with different
coupling parameters. We kept the coupling strength rather low to
ensure a clear separation of the time scales for Rabi oscillations and
for the pulses predicted in the previous section. We typically use a
magnetic field strength of $B_{\rm rf} = 0.1\,\mbox{mG}$ that is also
used in the latest experiments in the H\"ansch group \cite{BLOCH99:2}.
This amounts to a Rabi frequency of $\Omega = 2\pi \times
70\,\mbox{Hz}$, which is less than the trapping frequency $\omega_z$
and the pulse frequency $\delta_{\rm rf}$ (see below
Eq.~(\ref{eq:zmax})). All calculations were done with $N =
5\times10^5$ atoms in the trap.

\begin{figure}[!htbp]
  \begin{center}
    \epsfxsize=0.48\textwidth
    \epsffile{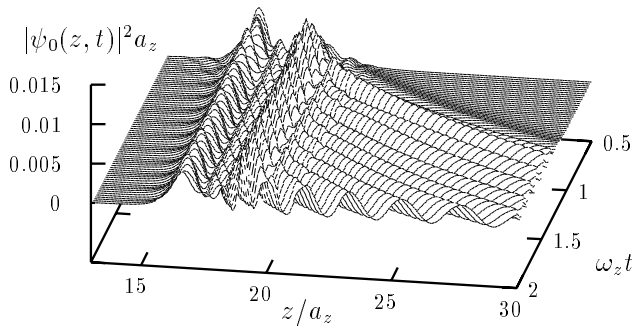}
    \caption{Plot of the outcoupled density.
      The atoms in the free $m=0$ state emerge primarily at the two
      resonance points (cf. Fig.~\ref{fig:densities2res}) and then
      fall downwards (to the right in the plot). One can clearly see
      the pulse trains emerging from the condensate region that
      extends to $z \approx 22\,a_z$. The radio frequency detunings
      in this plot are $\Delta_{1,2} = 2\pi\times (19.5\pm
      3.0) \,\mbox{kHz}$.}
    \label{fig:twomode3d}
  \end{center}
\end{figure}

\begin{figure}[tbp]
  \begin{center}
    \epsfxsize=0.47\textwidth
    \epsffile{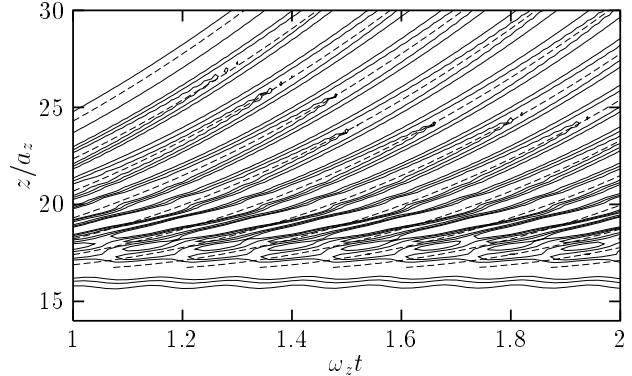}
    \caption{Contour plot of the data in Fig.~\ref{fig:twomode3d}. The 
      full lines are the contour lines for $|\psi_0(z,t)|^2 a_z =
      0.001,0.002,0.003$. The dashed lines denote the trajectories of the 
      pulse maxima from Eq.~(\ref{eq:zmax}).}
    \label{fig:contour}
  \end{center}
\end{figure}

In Fig.~(\ref{fig:twomode3d}) we show the density distribution of
the outcoupled state as a function of both space and time. The atomic
beam emerging from the condensate clearly exhibits the pulses predicted
in section \ref{sec:twowaves}.
To prove that these pulses really follow the trajectories in
Eq.~(\ref{eq:zmax}), Fig.~\ref{fig:contour} shows a contour plot of
the data in Fig.~\ref{fig:twomode3d}.  As we mentioned in
Sec.~\ref{sec:twowaves}, we used the phase $\alpha$ in
Eq.~(\ref{eq:zmax}) to shift the trajectories in time. Apart from
this, the pulses show the correct behavior, they start between the two
resonance points at $-\bar E/(Mg)$ and appear with the right
frequency.

\subsection{Output rates and visibility}
\label{sec:visib}

In this section we want to compare the numerical results to the
analytic predictions of Sec.~\ref{sec:airy}. All results were obtained
by keeping the frequency $\omegatwo$ fixed such that the corresponding
resonance point is at the center of the condensate ($\ztwo = z_{\rm
  sag} = g/\omegatrap^2, \Delta_2 = 2\pi\times 16.5\,\text{kHz}$).
$\Delta_1$ is then given by $\Delta_1 = \Delta_2 + \delta_{\rm
  rf}$; we consider only positive rf~differences $\delta_{\rm rf} = 0\ldots
2\pi\times 13\,\text{kHz}$.

\begin{figure}[!htbp]
  \begin{center}
    \epsfxsize=0.48\textwidth
    \epsffile{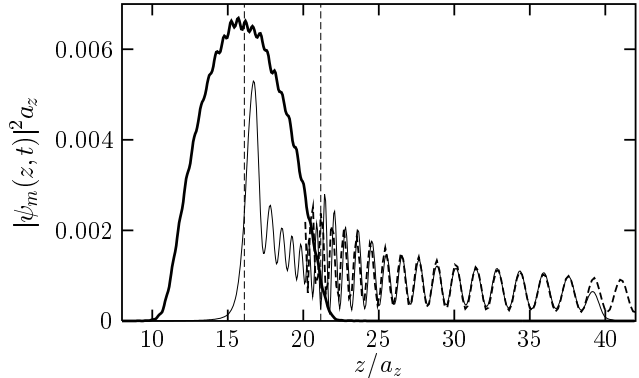}
    \caption{Plot of the density distributions of the atom laser (see
      also Fig.~\ref{fig:densities2res}). This plot is made after
      $5\,\text{ms}$ of operation with $\deltarf = 2\pi\times
      12000\,\text{Hz}$. The right resonance point is in the wing of
      the condensate so the visibility drops to $P=0.61$.
      The dashed line is a plot of the function in
      Eq.~(\ref{eq:psi02}) with the normalization of
      Eq.~(\ref{eq:norm}). The density of the untrapped state goes
      down to zero around $z=40\,a_z$ due to absorbing boundary
      conditions used in the calculations.}
    \label{fig:visibdensity}
  \end{center}
\end{figure}

The visibility in Fig.~\ref{fig:densities2res} is rather high, namely
$P = 0.90$. If one increases $\deltarf$, the visibility drops down, as
can be seen in Fig.~\ref{fig:visibdensity}. This is due to the fact
that it depends mainly on the quotient of the wavefunctions at the
resonance points (see
Eqs.~(\ref{eq:visib},\ref{eq:Pgamma},\ref{eq:gamma1d})), which
decreases if one pushes $\zone$ further to the boundary of the
condensate.
  
One can now take
Eqs.~(\ref{eq:gamma1d},\ref{eq:Pgamma},\ref{eq:psi02}) to calculate
the visibility and the norm of the Airy functions from
Sec.~\ref{sec:airy} using the values of $|\psi_{-1}(\zres)|$ either
from the Thomas-Fermi (TF) approximation \cite{DALF98:1} or directly
from the numerical calculations. The dashed line in
Fig.~\ref{fig:visibdensity} is obtained in this way using the
numerical values, outside the condensate it fits the numerical
calculations quite well. The Thomas-Fermi result (not shown) is only
slightly different. The total output rate $\Gamma^{\rm 1d}_{\zone} +
\Gamma^{\rm 1d}_{\ztwo}$ can be calculated via Eq.~(\ref{eq:norm})
from $\tilde{\cal N}^2$; we compared it to the initial decay constant
of the condensate occupation and found good agreement.

\begin{figure}[!htbp]
  \begin{center}
    \epsfxsize=0.48\textwidth
    \epsffile{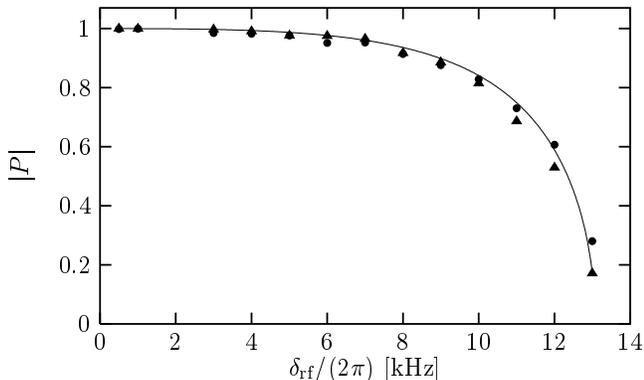}
    \caption{Visibility versus radio frequency difference. The full
      line is obtained from the Thomas-Fermi approximation. The
      triangles are calculated from the numerical values of
      $\psi_{-1}(\zres)$ at a specific time. The points denoted by
      circles are obtained by fitting the density distribution
      (\ref{eq:psi02}) to the numerical data at fixed time (after
      $\approx 7.5\,\text{ms}$ of outcoupling). The error for these
      values is smaller than 5\% and was conservatively deduced from
      the fitting procedure.}
    \label{fig:visibility}
  \end{center}
\end{figure}

After having shown that the output can be characterized by the
wavefunction in Eq.~(\ref{eq:psi02}), we want to compare further the
semi-analytic results of Sec.~\ref{sec:airy} to the numerics. For
different $\deltarf$ one can plot the visibilities obtained either by
fitting the data or by calculating $P$ via (\ref{eq:Pgamma}) using the
TF-approximation or the values of $|\psi_{-1}(\zres)|$ from the
numerics.  Fig.~\ref{fig:visibility} demonstrates the result: the
TF-approximation (full line) fits the semi-analytic values (triangles)
quite well. Like in an experiment, we tried to fit the density
distribution (\ref{eq:psi02}) to our numerical data at a fixed time to
get values for $P$. As the figure shows, these values are in a
reasonable agreement with the other results.
\begin{figure}[!htbp]
  \begin{center}
    \epsfxsize=0.48\textwidth
    \epsffile{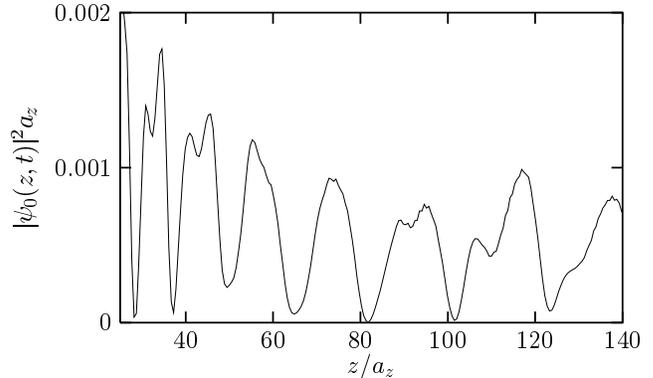}
    \caption{Density distribution of the untrapped state for $\deltarf
      = 2\pi\times 2000\,\text{Hz}$. The pulses have a substructure
      originating from roughly twice the difference frequency
      $\deltarf$.}
    \label{fig:pulses}
  \end{center}
\end{figure}
In Fig.~\ref{fig:visibility} we have not shown values for $\deltarf =
2\pi\times 2\,\mbox{kHz}$.  The reason for this is a resonance
phenomenon in this region that leads to a more complicated pulse
structure (Fig.~\ref{fig:pulses}) and prevents us from fitting
(\ref{eq:psi02}) to the data. The origin of this resonance is unclear
up to now, there are indications however that it is related to the
creation of small particle-like excitations in the condensate via a
process that is second order in the output coupling strength.

The small wiggles on top of the condensate density distribution in
Figs.~\ref{fig:densities2res} and \ref{fig:visibdensity} indicate the
presence of such excitations also for $\deltarf$ away from $2\pi\times
2\,\mbox{kHz}$. They are so small for these frequency differences that
the assumption made above Eq.~(\ref{eq:norm}) that the outcoupling
process proceeds as if the two radio frequencies were applied
separately is really satisfied, {\em i.e.} there are hardly any
processes of second order in the rf-coupling $\hbar\Omega$.

\section{Conclusion}
\label{sec:concl}

We have shown in this work that a rf-atom laser driven by two radio
frequencies --- a two-mode atom laser --- gives rise to a coherently
pulsing beam of atoms. We have given formulas for the output rate of
such a device and for the visibility of the pulse pattern of the
atomic beam. These analytic results are valid for atom lasers operated
at very low temperature (virtually $T=0$) and with weak output
coupling.

The main limitation of our results comes from dimensionality: we use a
1D model for our calculations instead of a full 3D treatment. In
\cite{SCHNEI99:1} we have shown that the output rate of a rf-atom
laser operated with one frequency can be calculated by the 1D model in
qualitative agreement with experiment. Accordingly, apart from
quantitative agreement of the output rates our model reproduces the
experimental features quite well \cite{BLOCH99:2}.

We did not address the problem of a pump for a true cw-atom laser.
Instead, we considered rather small couplings that ensure a continuous
and smooth operation for about $20-50\,\text{ms}$ but with slowly
decreasing output rates. After this time only very few atoms are left
in the atom trap. The output of a true cw-atom laser should
be very similar to that of our situation.

Coherently pulsing beams of atoms are certainly of interest for
applications in quantum and atom optics. The setup described in this
paper produces beams that always point downwards due to gravity. It
should be possible to use a similar setup with two different
two-photon Raman transitions to create a pulsed beam that can be
pointed in any direction by using the momentum kick from the
absorption-emission process.

Finally, if one uses more than two frequencies, one might create
pulsed beams with shorter pulses while keeping their repetition rate
fixed in analogy to a mode-locked optical laser \cite{ANDE98:1}.

\acknowledgements
We thank Immanuel Bloch and Tilman Esslinger for fruitful discussions.
Financial support by DFG under Grant Nr. SCHE~128/7-1 is
gratefully acknowledged.

%\bibliographystyle{prsty}
%\bibliography{lit}

%\begin{thebibliography}{10}

%\include{loglabs}

\end{document}